\documentclass[letterpaper, 10 pt, conference]{class/ieeeconf}  

\IEEEoverridecommandlockouts                              
\usepackage{amsmath}
\usepackage{amssymb}
\usepackage{latexsym}
\usepackage{url}
\usepackage{cite}
\usepackage{relsize}
\usepackage{multirow}
\usepackage{afterpage}
\usepackage{ifthen}
\usepackage{graphicx}
\usepackage{algpseudocode,algorithm,algorithmicx}
\usepackage{subfigure}
\usepackage[keeplastbox]{flushend}
\usepackage{epstopdf}
\usepackage{stackengine}
\usepackage{gensymb}
\usepackage{booktabs}
\usepackage{makecell}
\usepackage{hhline}
\usepackage{courier}
\usepackage{lipsum}
\usepackage{xcolor}
\usepackage{bm}

\title{\LARGE \bf End-to-End Real-time Catheter Segmentation with Optical Flow-Guided Warping during Endovascular Intervention}

\author{Anh Nguyen$^1$, Dennis Kundrat$^1$, Giulio Dagnino$^1$, Wenqiang Chi$^1$, Mohamed E. M. K. Abdelaziz$^1$, \\ Yao Guo$^1$, YingLiang Ma$^2$, Trevor M. Y. Kwok$^3$, Celia Riga$^3$, and Guang-Zhong Yang$^{1,4}$
\thanks{$^1$Hamlyn Centre for Robotic Surgery, Imperial College London, UK {\tt a.nguyen@imperial.ac.uk}}
\thanks{$^2$School of Computing, Electronics and Mathematics, Coventry University, UK}
\thanks{$^3$Department of Surgery and Cancer, Imperial College London, St Mary's Hospital, UK}
\thanks{$^4$Institute of Medical Robotics, Shanghai JiaoTong University, China}
}
\begin{document}

\newtheorem{problem}{Problem}
\newtheorem{lemma}{Lemma}
\newtheorem{theorem}[lemma]{Theorem}
\newtheorem{claim}{Claim}
\newtheorem{corollary}[lemma]{Corollary}
\newtheorem{definition}[lemma]{Definition}
\newtheorem{proposition}[lemma]{Proposition}
\newtheorem{remark}[lemma]{Remark}
\newenvironment{LabeledProof}[1]{\noindent{\it Proof of #1: }}{\qed}

\def\beq#1\eeq{\begin{equation}#1\end{equation}}
\def\bea#1\eea{\begin{align}#1\end{align}}
\def\beg#1\eeg{\begin{gather}#1\end{gather}}
\def\beqs#1\eeqs{\begin{equation*}#1\end{equation*}}
\def\beas#1\eeas{\begin{align*}#1\end{align*}}
\def\begs#1\eegs{\begin{gather*}#1\end{gather*}}

\newcommand{\poly}{\mathrm{poly}}
\newcommand{\eps}{\epsilon}
\newcommand{\e}{\epsilon}
\newcommand{\polylog}{\mathrm{polylog}}
\newcommand{\rob}[1]{\left( #1 \right)} 
\newcommand{\sqb}[1]{\left[ #1 \right]} 
\newcommand{\cub}[1]{\left\{ #1 \right\} } 
\newcommand{\rb}[1]{\left( #1 \right)} 
\newcommand{\abs}[1]{\left| #1 \right|} 
\newcommand{\zo}{\{0, 1\}}
\newcommand{\zonzo}{\zo^n \to \zo}
\newcommand{\zokzo}{\zo^k \to \zo}
\newcommand{\zot}{\{0,1,2\}}
\newcommand{\en}[1]{\marginpar{\textbf{#1}}}
\newcommand{\efn}[1]{\footnote{\textbf{#1}}}
\newcommand{\vecbm}[1]{\boldmath{#1}} 
\newcommand{\uvec}[1]{\hat{\vec{#1}}}
\newcommand{\thv}{\vecbm{\theta}}
\newcommand{\junk}[1]{}
\newcommand{\var}{\mathop{\mathrm{var}}}
\newcommand{\rank}{\mathop{\mathrm{rank}}}
\newcommand{\diag}{\mathop{\mathrm{diag}}}
\newcommand{\tr}{\mathop{\mathrm{tr}}}
\newcommand{\acos}{\mathop{\mathrm{acos}}}
\newcommand{\atantwo}{\mathop{\mathrm{atan2}}}
\newcommand{\SVD}{\mathop{\mathrm{SVD}}}
\newcommand{\quadf}{\mathop{\mathrm{q}}}
\newcommand{\linterp}{\mathop{\mathrm{l}}}
\newcommand{\sgn}{\mathop{\mathrm{sign}}}
\newcommand{\sym}{\mathop{\mathrm{sym}}}
\newcommand{\avg}{\mathop{\mathrm{avg}}}
\newcommand{\mean}{\mathop{\mathrm{mean}}}
\newcommand{\erf}{\mathop{\mathrm{erf}}}
\newcommand{\grad}{\nabla}
\newcommand{\R}{\mathbb{R}}
\newcommand{\defeq}{\triangleq}
\newcommand{\dims}[2]{[#1\!\times\!#2]}
\newcommand{\sdims}[2]{\mathsmaller{#1\!\times\!#2}}
\newcommand{\udims}[3]{#1}
\newcommand{\udimst}[4]{#1}
\newcommand{\com}[1]{\rhd\text{\emph{#1}}}
\newcommand{\ind}{\hspace{1em}}
\newcommand{\argmin}[1]{\underset{#1}{\operatorname{argmin}}}
\newcommand{\floor}[1]{\left\lfloor{#1}\right\rfloor}
\newcommand{\step}[1]{\vspace{0.5em}\noindent{#1}}
\newcommand{\quat}[1]{\ensuremath{\mathring{\mathbf{#1}}}}
\newcommand{\norm}[1]{\left\lVert#1\right\rVert}
\newcommand{\ignore}[1]{}
\newcommand{\specialcell}[2][c]{\begin{tabular}[#1]{@{}c@{}}#2\end{tabular}}
\newcommand*\Let[2]{\State #1 $\gets$ #2}
\newcommand{\algorithmicbreak}{\textbf{break}}
\newcommand{\Break}{\State \algorithmicbreak}
\newcommand{\ra}[1]{\renewcommand{\arraystretch}{#1}}

\renewcommand{\vec}[1]{\mathbf{#1}} 

\algdef{S}[FOR]{ForEach}[1]{\algorithmicforeach\ #1\ \algorithmicdo}
\algnewcommand\algorithmicforeach{\textbf{for each}}
\algrenewcommand\algorithmicrequire{\textbf{Require:}}
\algrenewcommand\algorithmicensure{\textbf{Ensure:}}
\algnewcommand\algorithmicinput{\textbf{Input:}}
\algnewcommand\INPUT{\item[\algorithmicinput]}
\algnewcommand\algorithmicoutput{\textbf{Output:}}
\algnewcommand\OUTPUT{\item[\algorithmicoutput]}

\maketitle
\thispagestyle{empty}
\pagestyle{empty}


  


\begin{abstract}

Accurate real-time catheter segmentation is an important pre-requisite for robot-assisted endovascular intervention. Most of the existing learning-based methods for catheter segmentation and tracking are only trained on small-scale datasets or synthetic data due to the difficulties of ground-truth annotation. Furthermore, the temporal continuity in intraoperative imaging sequences is not fully utilised. In this paper, we present FW-Net, an end-to-end and real-time deep learning framework for endovascular intervention. The proposed FW-Net has three modules: a segmentation network with encoder-decoder architecture, a flow network to extract optical flow information, and a novel flow-guided warping function to learn the frame-to-frame temporal continuity. We show that by effectively learning temporal continuity, the network can successfully segment and track the catheters in real-time sequences using only raw ground-truth for training. Detailed validation results confirm that our FW-Net outperforms state-of-the-art techniques while achieving real-time performance.

\end{abstract}

\section{INTRODUCTION} \label{Sec:Intro}
In cardiovascular surgery, endovascular intervention offers many advantages compared to the traditional open surgical approaches, including smaller incisions, less trauma for patients, local instead of general anesthesia, stability, and more importantly, reduced risks for patients who have comorbidities~\cite{simaan2018medical}. Endovascular intervention involves the manipulation of catheters and guidewires to reach target areas in the vasculature to deliver a treatment (e.g. stenting, ablation or drug delivery~\cite{rafii-tari_current_2014}). Such tasks require a high level of technical skills to avoid damage to the vessel wall, which could result in perforation and hemorrhage, or dissection and organ failure, all of which can be fatal. Despite their relative advantages, endovascular procedures still present some limitations such as limited sensory feedback, misalignment of visuo-motor axes, and the need for high dexterity from the operators~\cite{benavente_molinero_haptic_nodate}. Robotics and computer assistance have been integrated into the clinical workflow to provide augmentation of surgical skills in terms of enhanced dexterity and precision~\cite{thakur2009design, abdelaziz2019toward, bian2013enhanced, bobby20, zhao2019cnn, varghese2020nonlinearity, dagnino2018haptic, chi2018learning, Cursi2020HybridDA, zhu2017deep}.

As a pre-requisite to robot-assisted intervention, the task of catheter segmentation can provide essential visual or haptic feedback for the surgeons. For example, in~\cite{dagnino2018haptic}, a vision-based force sensing is developed based on the tip position of catheter and the vasculature. However, in routine practice, damages to the vessels are generated not only by the contact of the catheter tip with the vessel wall, but by the contacts between the entire catheter and the endothelial wall. Therefore, delineation and tracking of the entire catheter are essential. However, autonomous catheter segmentation is not a trivial task for two main reasons. Firstly, in the X-ray image, catheters can be easily confused with other similar linear structures like blood vessels due to its low contrast. Secondly, during clinical trials, catheters and guidewires can have a sudden and large deformation movement. This leads to the fact that traditional methods~\cite{brost2009_tracking, brost2010respiratory, yatziv2012_multiple, baert2003_guidewire_tracking} based on primitive features of catheter appearance have limited generality and would not be able to segment catheters in real-time and dynamic surgical environments.

\begin{figure}[!t] 
    \centering
    \includegraphics[width=0.99\linewidth, height=0.8\linewidth]{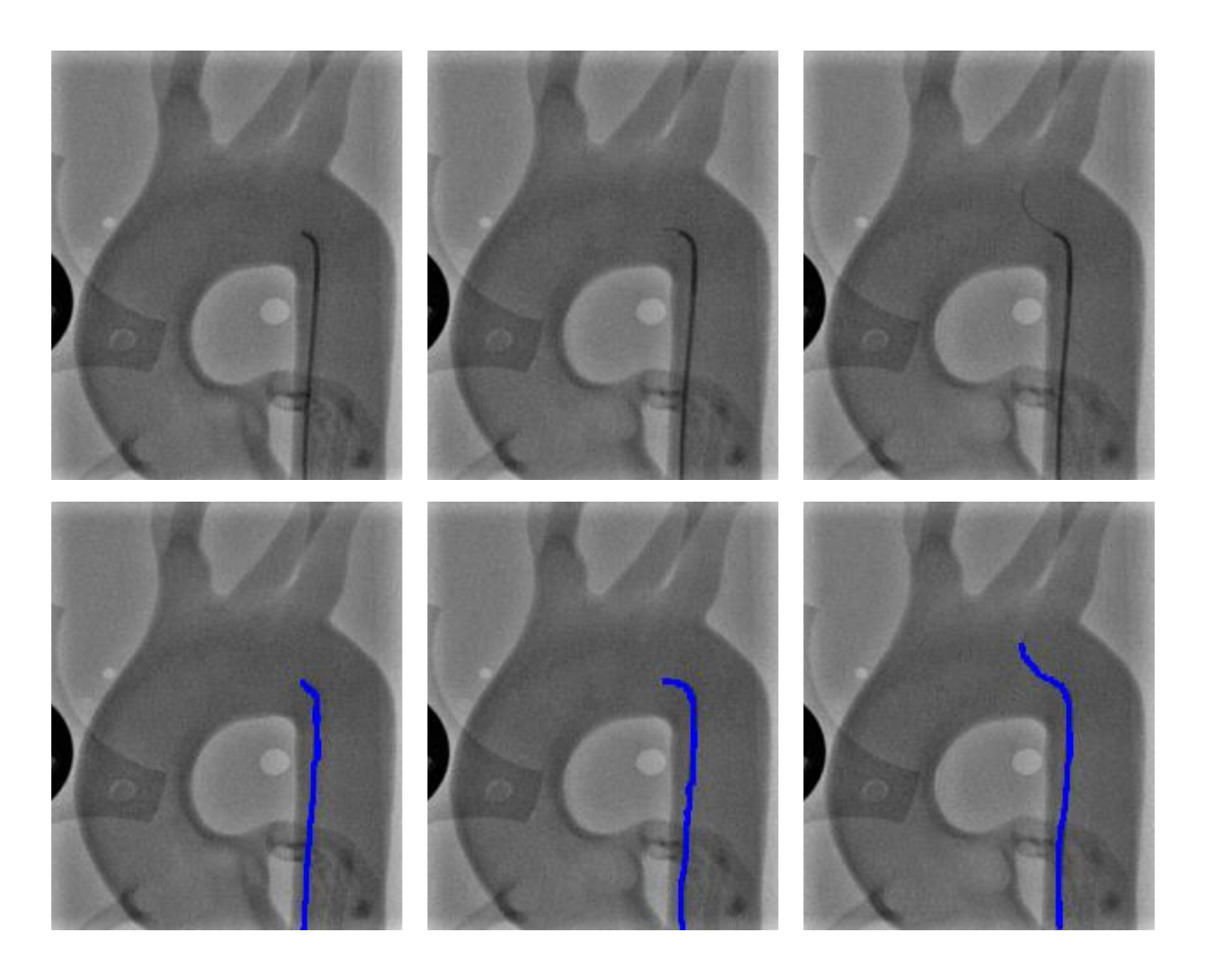} 
    \vspace{0.2ex}
    \caption{Catheter segmentation in 2D X-ray fluoroscopy sequences. \textbf{Top row:} The X-ray images of the catheter advancing within an aortic phantom. \textbf{Bottom row:} An illustration of segmented results.}
    \label{Fig_intro} 
\end{figure}

Recently, machine learning, especially deep learning has been widely adopted as a novel approach for medical image segmentation~\cite{ronneberger2015_UNet, milletari2016_VNet, wang2018deepigeos}. The effectiveness of deep learning comes from the ability to handle a large amount of multimodal input data~\cite{nguyen2019objectC, lecun2015deep, nguyen2019v2cnet}. However, this advantage becomes a potential problem in catheter segmentation since it is not easy to create large-scale datasets with pixel-wise labels. This is because the annotation task requires a certain amount of medical expertise, while manually labeling is very tedious, especially for objects with elongated structures such as catheters and guidewires. Due to these challenges, recent deep learning methods for catheter segmentation mainly train on a very small dataset~\cite{breininger2018_multihead}~\cite{breininger2018intraoperative}, use synthetic data~\cite{yi2019_rnn_synthetic}, or create ground-truth based on a particular observation about pixel intensity~\cite{vlontzos2018deep}. These assumptions technically limit the power of deep learning and the generality of the methods.

In this paper, we propose Flow-guided Warping Net (FW-Net), a new end-to-end framework for catheter segmentation in 2D X-ray fluoroscopy sequences (Fig.~\ref{Fig_intro}). Our hypothesis is that deep network can be trained using the raw ground-truth while the overall accuracy can be improved by effectively learning the temporal continuity from X-ray sequences. In particular, we first create the raw ground-truth using a vision-based approach~\cite{ma2018novel}, then we design FW-Net with three modules: \textit{i)} a segmentation network, \textit{ii)} a flow network, and \textit{iii)} a novel flow-guided warping function. We train FW-Net on raw ground-truth data and employ the flow-guided warping function to learn the temporal continuity between consecutive X-ray frames. This will encourage the network to predict based on both the raw ground-truth and sequential information, hence potentially improve the accuracy.

The rest of the paper is organized as follows. We review the related work in Section~\ref{Sec_rw}, then describe the data collection process on our robotic platform in Section~\ref{Sec_robot}. In Section~\ref{Sec_method}, we present the new end-to-end architecture for effectively segmenting the catheter from raw ground-truth. The experimental results are presented in Section~\ref{Sec_exp}. Finally, we conclude the paper and discuss the future work in Section~\ref{Sec_con}.

\section{Related Work} \label{Sec_rw}
Recently, there has been an increasing effort in segmenting catheters and guidewires from X-ray images. These methodologies can be divided into two main categories: vision-based approach and learning-based approach.

Traditional methods for catheter segmentation mainly used primitive image level cues such as pixel intensity, texture, or histogram~\cite{brost2009_tracking, brost2010respiratory, wu2011learning, yatziv2012_multiple, ma2013real, baert2003_guidewire_tracking}. In~\cite{sheng2009automatic}, the authors introduced a method based on Hough transform for detecting supporting device position in adult chest X-ray. Similarly, Kao et al. ~\cite{kao2015_chest_CTT} proposed a system to detect endotracheal tubes on pediatric chest X-ray image using local features and multiple thresholds. Keller et al.\cite{keller2007semi} introduced a semi-automated method for catheter detection and tracking using prior information from users input. Mercan et al.  \cite{bismuth2012curvilinear} proposed to use local and global curvature features with controllable smoothness for guidewire segmentation. More recently, the authors in~\cite{ma2018novel} used the multiscale vessel enhancement filter and adaptive binarization technique for detecting catheters and guidewires in real-time. A major drawback of all methods based on thresholding techniques is they do not generalize well and are very sensitive to a particular input X-ray data. 

Machine learning techniques are also widely used for catheter segmentation and tracking~\cite{wang2009_tracking, chen2016guidewire, wang2017_detection, pauly2010machine}. With the rise of deep learning, methods based on Convolutional Neural Networks (CNN) are adapted for catheter segmentation~\cite{yang2019improving}~\cite{zaffino2019fully}. Early work in \cite{mercan2013_chesttube} used a simple neural network to detect chest tubes then post-processed the results using a curve fitting technique to connect discontinued segments. In~\cite{ronneberger2015_UNet}\cite{milletari2016_VNet}, the state-of-the-art U-Net and V-Net architecture were introduced for data-driven medical image segmentation. Ambrosini et al.~\cite{ambrosini2017fully} presented an adaptive U-Net architecture for catheter segmentation in X-ray sequences. Vlontzos and Mikolajczyk~\cite{vlontzos2018deep} segmented the catheter from X-ray angiography video with a deep network and the ground-truth created by a carefully manual thresholding. Unberath et al.~\cite{unberath2018_mri_xray} presented a framework for simulating fluoroscopy and digital radiography from CT scans, then detecting anatomical landmarks with a deep network. The authors in~\cite{breininger2018_multihead}~\cite{breininger2018intraoperative} used CNN with multihead for stent segmentation in X-ray fluoroscopy images. More recently, in~\cite{yi2019_rnn_synthetic} a scale-recurrent network was used to detect catheters in synthetic X-ray data.

While deep learning-based approaches can learn meaningful features from input data, applying deep learning to catheter segmentation problem is not straightforward due to the lack of real X-ray data, and the tediousness when manually labeling ground-truth. In this work, we propose to learn from raw ground-truth data and encode the temporal consistency between neighborhood X-ray frames. This will help the network rely more on the temporal information to segment the catheter in X-ray sequences. 
\section{Data Collection} \label{Sec_robot}

\begin{figure}[h] 
    \centering
    \includegraphics[width=0.99\linewidth, height=0.55\linewidth]{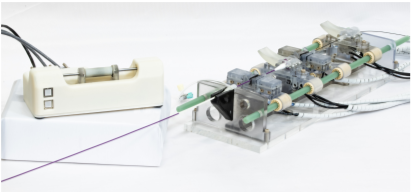} 
    \vspace{1ex}
    \caption{CathBot robotic platform for fluoroscopy and MR-guided endovascular interventions. \textbf{Left:} Master device. \textbf{Right:} MR-safe slave robot. }
    \label{Fig_catbot} 
\end{figure}

\begin{figure*}[!ht] 
    \centering
    \includegraphics[width=0.99\linewidth, height=0.5\linewidth]{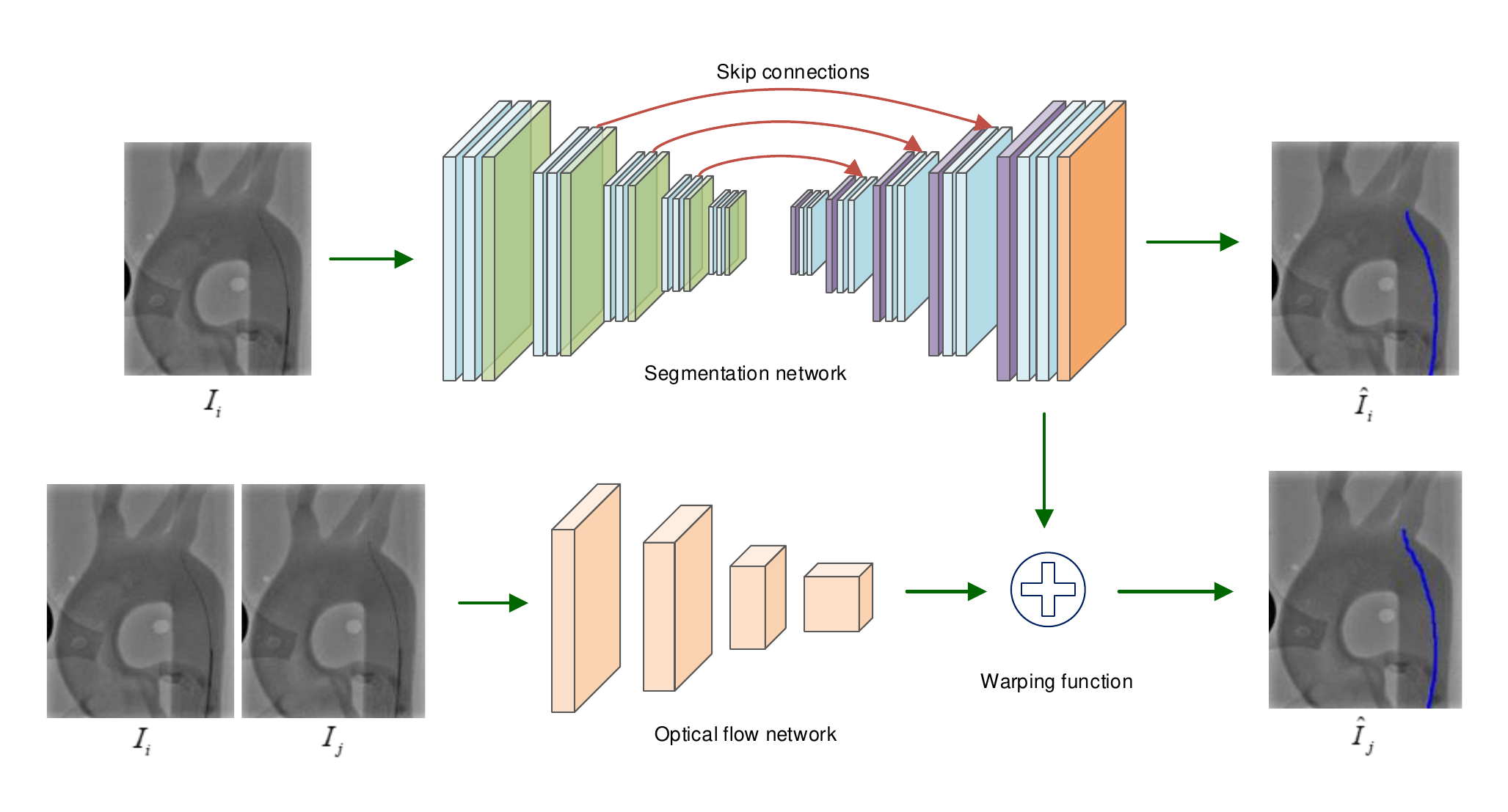} 
    \vspace{1ex}
    \caption{An overview of our FW-Net architecture. The network consists of three modules: a segmentation network with encoder-decoder architecture and skip connections, a flow network to extract optical flow information from two neighborhood frames, and a flow-guided warping function to learn the frame-to-frame temporal continuity.}
    \label{Fig_method} 
\end{figure*}

\subsection{CathBot}

In this work, we collect sequences of X-ray data during the intervention using the CathBot \cite{abdelaziz2019toward} robot. CathBot (Fig.~\ref{Fig_catbot}) comprises a versatile master-slave setup and navigation framework. Unlike previous platforms, the robot can be safely integrated and used in Magnetic Resonance (MR) environments thanks to pneumatic actuation and additive manufacturing. The master robot is an intuitive human machine interface (HMI) which mimics the human motion pattern (i.e. grasping the instrument followed by insertion/retraction and/or rotation) and provides haptic feedback to the users generated by the navigation systems as described in~\cite{benavente_molinero_haptic_nodate, dagnino2018haptic}. Motions are mapped to the 4-DOF MR-safe slave robot, capable of manipulating off-the-shelf catheters and guidewires.

\subsection{X-ray Data Collection}
A vascular soft silicone phantom (Elastrat, Geneva, Switzerland) of a normal adult human aortic arch was placed underneath an X-ray imaging system to simulate a patient lying on the angiography table to undergo an endovascular procedure. The phantom was connected to a pulsatile pump to simulate normal human blood flow and optimize the level of realism for tool-tissue interactions. A professional surgeon was asked to cannulate three arteries by manipulation of wire and catheter. Namely, the left subclavian (LSA), left common carotid (LCCA) and right common carotid (RCCA) arteries. The cannulation was performed in two scenarios: manual and robot assisted. During each maneuver, fluoroscopy was activated by the operator using a pedal. Real-time video stream of the surgical scene was acquired using an image grabber (DVI2USB3, Epiphan Video, Ottawa, Canada) from a vascular imaging system - in this study we have used a fluoroscopic system for interventional radiology procedure (Innova 4100 IQ GE Healthcare). The video stream was acquired on a workstation (Windows 7, Intel i7-6700, 3.4GHz, 16GB RAM) and digitalized into image sequence for image processing.

\section{Methodology} \label{Sec_method}
Our goal is to segment the catheters and guidewires in X-ray fluoroscopy sequences using the raw ground-truth created by~\cite{ma2018novel}. Since the selected ground-truth annotation method does not take into account the temporal continuity, which is the key information from the X-ray sequences, we construct a unified framework to effectively learn this information. Towards this end, we propose FW-Net, a new end-to-end architecture to effectively segment the catheter in X-ray sequences using a novel flow-guided warping function. The overall architecture of our proposed approach is illustrated in Fig.~\ref{Fig_method}.

\subsection{Segmentation Network}
Our specific segmentation task is to compute a binary mask separating the foreground (i.e., catheter and guidewire) from the background for every X-ray frame of the video. Inspired by the effectiveness of deep neural networks in image segmentation, we build our segmentation branch based on encoder-decoder architecture~\cite{ronneberger2015_UNet}~\cite{nguyen2016detecting}. To improve the real-time performance of the network, we use big convolution kernels with large strides to extract features from the input X-ray frame. Since the convolution operation is comparably cheap with a small number of channels as in X-ray images, using big kernels does not significantly increase the computational costs. Furthermore, we combine large strides with skip connections as in U-Net architecture~\cite{ronneberger2015_UNet} to maintain low-level features during the decoding process.

Specifically, the input of the segmentation network is the RGB X-ray image of size ($256 \times 256$) pixels. The encoder network has $5$ ResNet blocks~\cite{he2016deep} to extract the depth features from input images. Each ResNet block consists of a convolutional layer, ReLU, skip links and pooling operations. The output map after each ResNet block in the encoder network has the size of $128, 64, 32, 16$, and $8$ respectively. Each decoder block is associated with an encoder. In each decoder block, the encoder feature map is upsampled using the deconvolutional operation. Finally, a $2$ classes soft-max layer is used at the end of the decoder network to classify the background and foreground for all pixels in the current X-ray frame.

Unlike the traditional image segmentation problem, in catheter segmentation, the imbalance between the foreground and background regions is strongly significant since the foreground only occupies a small portion number of pixels. To overcome this problem, we employ the weighted version of pixel-wise cross-entropy loss function $\mathbb{E}$ as in~\cite{xie2015holistically}. The segmentation loss is defined as follows:

\begin{equation}
\begin{aligned}
\mathcal{L}_s = -(1 - w)\sum\limits_{i,j \in \mathrm{fg}} {\mathop{\rm \log\mathbb{E}}\nolimits} ({y_{ij}} = 1;\theta ) \\ - w \sum\limits_{i,j \in \mathrm{bg}} {{\mathop{\rm \log\mathbb{E}}\nolimits} ({y_{ij}} = 0;\theta )}
 \end{aligned}
\end{equation}
where $i$ and $j$ are the pixel location of the foreground $\mathrm{fg}$ and the background $\mathrm{bg}$, respectively; $y_{ij}$ denotes the binary prediction of each pixel of the input image, $w$ is the foreground-background pixel-number ratio, and $\theta$ represents the network parameters.

\subsection{Optical Flow Network}
Extracting optical flow is a fundamental task in video analysis. Traditional methodologies for this problem have been studied for decades and mainly used variational approaches which address small displacements~\cite{weickert2006survey}. Recently, deep learning has been exploited for learning optical flow. In this work, we adopt the simple version of FlowNet~\cite{dosovitskiy2015flownet}, a state-of-the-art deep neural-based architecture as our flow network. To decrease the computational complexity, we reduce the number of convolutional kernels in each layer of FlowNet by half and hence reduce the overall complexity to one fourth.

In practice, we stack two neighborhood X-ray images ($I_i, I_j$) together and feed them through a deep network to extract the flow motion. Note that, the $I_i$ frame is also the input frame for segmentation network. Since the computed optical flow is aligned with the segmentation output, their shared feature map information can be combined later naturally to generate the segmented map for $I_j$. Specifically, our flow network has a sequence of $6$ convolutional layers to estimate the flow motion from consecutive video frames. All convolutional layers have the stride of $2$. Compared to the segmentation network, the flow network is simpler with fewer parameters. 

\subsection{Flow-Guided Warping Function}
 Unlike the traditional image segmentation problem, where the temporal information is not available, in video segmentation, temporal consistency across frames is the key to success. Our observation is that the consecutive X-ray frames are highly similar. This similarity is even stronger in the deep feature maps since they encode high level semantic concepts from these frames~\cite{jayaraman2016slow}. We exploit the similarity by warping the deep features from segmentation network with the flow motion from flow network.

As motivated by Zhu et al.~\cite{zhu2017deep}, given a reference frame $I_i$ and a neighbor frame $I_j$, a flow motion field ${\mathrm{M}_{i \to j}} = \mathcal{F}({I_i},{I_j})$ is estimated by a flow network $\mathcal{F}$ (e.g., FlowNet). The feature maps on the reference frame are warped to the neighbor frame according to the optical flow. The warping function is defined as:

\begin{equation}
{f_{i \to j}} = \mathcal{W}\left( {{f_i},{\mathrm{M}_{i \to j}}} \right) = \mathcal{W}\left( {{f_i},\mathcal{F}({I_i},{I_j}} \right))
\end{equation}
where ${f_{i \to j}}$ denotes the feature maps warped from previous frame $I_i$ to frame $I_j$. $\mathcal{W}(\cdot)$ is the bilinear warping function applied on all the locations for each channel in the feature maps. $\mathcal{F}({I_i},{I_j})$ is the flow field estimated by the flow network, which maps a location $\mathbf{p}=(p_x, p_y)$ in frame $I_i$ to the location $\mathbf{p}+\delta \mathbf{p}$ in frame $I_j$.

Since the feature maps has several channels, the warping is performed in each channel as:

\begin{equation}
{\varphi _{i \to j}}(\mathbf{p}) = \sum\limits_\mathbf{q} {\mathbf{K}(\mathbf{q},\mathbf{p} + \delta \mathbf{p}){\varphi _j}(\mathbf{q})}
\end{equation}
where $\mathbf{q}$ denotes all spatial locations in the feature maps, and $\mathbf{K}$ indicates the bilinear interpolation kernel.

Since we employ end-to-end training, the backprogagation of ${\varphi _{i \to j}}$ with respect to $\varphi_i$ and flow $\delta \mathbf{p}$ is derived as:

\begin{equation}
\begin{aligned}
\frac{{\partial {\varphi _{i \to j}}(\mathbf{p})}}{{\partial {\varphi _i}(\mathbf{q})}} &= \mathbf{K}(\mathbf{q},\mathbf{p} + \delta \mathbf{p})\\
\frac{{\partial {\varphi _{i \to j}}(\mathbf{p})}}{{\partial \mathcal{F}({I_i},{I_j})(\mathbf{p})}} &= \sum\limits_\mathbf{q} {\frac{{\partial \mathbf{K}(\mathbf{q},\mathbf{p} + \delta \mathbf{p})}}{{\partial \delta \mathbf{p}}}} {\varphi _i}(\mathbf{q})
\end{aligned}
\end{equation}

Intuitively, the warping function $\mathcal{W}(\cdot)$ combines the features of the segmentation network with the output of the flow network in the same region of the reference frame $I_i$, then generate the segmentation for that region in the neighbor frame $I_j$. This warping process provides more diverse information on the same image region, such as deformation and varied illuminations while effectively use the temporal information from the flow. We also note that the flow network cannot generate the semantic segmentation by itself since it only predicts the displacement by optical flow. Therefore, we need to combine the flow network with the segmentation network using the warping function $\mathcal{W}(\cdot)$ to generate the segmentation map for the neighbor frame.

\textbf{Training}
 The network is end-to-end trained using stochastic gradient descent (SGD) with a fixed $0.001$ learning rate and $0.9$ momentum. In each mini-batch, a pair of nearby video frames ($I_i, I_j$) with $0 \le j - i \le 6$, are randomly sampled. The total loss is the combination of two cross-entropy losses as follows:
 \begin{equation}
     \mathcal{L} = \mathcal{L}_s + \lambda \mathcal{L}_w
 \end{equation}
 where $\mathcal{L}_s$ is the loss of segmentation network to generate segmented map for $I_i$, and $L_w$ is the loss for generating segmented map for $I_j$. $\lambda$ is the hyperparameter and is empirically set to $0.4$.
 
 In practice, we implement our method using the Tensorflow library~\cite{abadi2016tensorflow}. The network is trained from scratch until convergence with no further reduction in training loss. The training time is approximately 2 days on an NVIDIA GTX 2080 GPU on a dataset with more than $20.000$ X-ray frames.
\section{Experiments} \label{Sec_exp}
\subsection{Experimental Setup}
\textbf{Dataset} 
We perform $28$ clinical trials using the CathBot robot, resulting in $28$ X-ray videos. Each video describes the movement of the catheter and guidewire in each trial and is approximately $2 - 5$ minutes long. We extract X-ray frames from each video at $8$ frames per second. In total, our new X-ray dataset has $25,271$ frames from $28$ sequences. We resize all the frames to $(256 \times 256)$ pixels before using them in our network. The raw pixel-wise ground-truth of the frames is created using the method in~\cite{ma2018novel}. Due to the nature of the technique in~\cite{ma2018novel}, both the catheter and guidewire are considered as one class in our experiment. For quantitative evaluation, we manually label $2$ sequences with approximately $1,000$ frames for testing, and use all frames from the other videos for training. We notice all labels for training are created automatically by~\cite{ma2018novel}, and no further manual human correction is needed.

\textbf{Metric}
As the standard practice in binary segmentation, we use the $Dice$ metric to evaluate the segmentation results. The $Dice$ index is defined between the ground-truth mask $X$ and the predicted mask $Y$ as follows:

\begin{equation}
  Dice = \frac{{2||X \cap Y||}}{{||X|| + ||Y||}} = \frac{{2TP}}{{2TP + FP + FN}}  
\end{equation}
where $TP$ denotes the true positive number of labeled pixels, $FP$ indicates the false positive pixels, and $FN$ is the false negative pixels.

\textbf{Baseline}
We compare our results (FW-Net) with the following state-of-the-art methods: U-Net~\cite{ronneberger2015_UNet}, Siamese U-Net~\cite{vlontzos2018deep}, Adaptive U-Net~\cite{ambrosini2017fully}, and TCF~\cite{ma2018novel}. We note that TCF~\cite{ma2018novel} does not require training since it used vision-based technique, while all other methods are trained on the same training set with raw ground-truth. Within the deep learning methods, the U-Net architecture does not take into account the sequential information, while our network, Siamese U-Net~\cite{vlontzos2018deep}, and Adaptive U-Net~\cite{ambrosini2017fully} exploit the use of temporal information.

\subsection{Results}

\begin{table}
\centering\ra{1.3}
\caption{Dice Scores over the Testing Set}
\label{tb_result}
\hspace{2ex}

\begin{tabular}{@{}rcccccccc@{}}
\toprule 					&  
Training?     & 
Temporal?     & 
FPS           &
$Dice$          & 

\\
\midrule
TCF~\cite{ma2018novel} 	& No    & No   &  10 (CPU)  & 0.796    		  \\
U-Net~\cite{ronneberger2015_UNet} 		    & Yes   & No   &  2 (GPU)	 & 0.677  		   \\
Adaptive U-Net~\cite{ambrosini2017fully}   & Yes   & Yes  & 8 (GPU)	 & 0.745    \\
Siamese U-Net~\cite{vlontzos2018deep}  	& Yes   & Yes  & 90 (GPU) & 0.768		   \\
\cline{1-6}
FW-Net (ours)			        & Yes   & Yes  & 15 (GPU) & \textbf{0.821} \\		
\bottomrule
\end{tabular}
\end{table}

\begin{figure*}[!ht] 
    \centering
    \includegraphics[width=0.99\linewidth, height=0.97\linewidth]{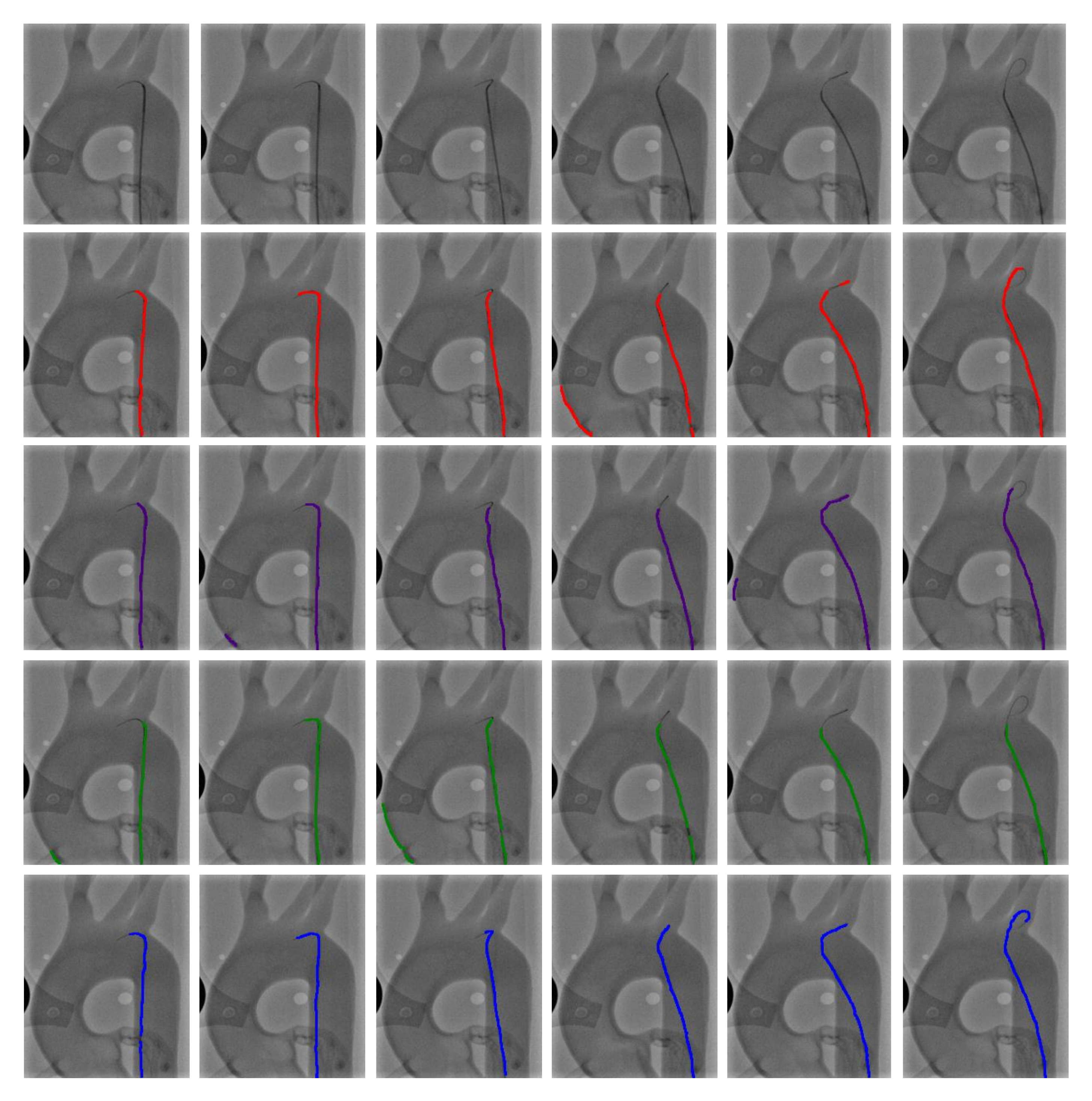} 
    \vspace{1ex}
    \caption{A visualization of the segmentation results of an X-ray image sequence. Top row - original X-ray images; Second row - TCF~\cite{ma2018novel}; Third row: Adaptive U-Net~\cite{ambrosini2017fully}; Fourth row: Siamese U-Net~\cite{vlontzos2018deep}; Fifth row: Our FW-Net. Compared to other methods, our FW-Net shows less over-segmented regions as well as less disconnected segments.  }
    \label{Fig_result} 
\end{figure*}

Table~\ref{tb_result} summarizes segmentation results of our method and the baselines on the testing set. The results clearly show that our FW-Net consistently improves over the state-of-the-art. In particular, FW-Net achieves the $Dice$ score of $0.821$, which is a concrete improvement over the second-best method. It is worth noting that FW-Net outperforms the original approach U-Net approach by a large margin of $14.3\%$. This result is explainable since the U-Net architecture is designed for an individual frame and does not take into account the temporal information, while our FW-Net is designed to learn the frame-to-frame temporal continuity effectively from X-ray sequences.  

We also observe a significant improvement of our FW-Net over Adaptive U-Net and Siamese U-Net, which are the deep learning-based methods exploit the temporal information. It shows that our proposed flow-guided warping method can encode the temporal information more successfully than Adaptive U-Net (which only trains the video frame sequentially) or Siamese U-Net (which relies heavily on data augmentation). We also found that all networks exploit temporal information achieve better results than the original U-Net. However, since all the network are trained using the raw ground-truth, other deep networks except our FW-Net cannot outperform the classical TCF method. 

Table~\ref{tb_result} also provides the intuitive inference time of all methods in frame per second (FPS). Overall, our FW-Net achieves a speed of $15$ FPS on NVIDIA GTX 2080 GPU, which is reasonable for real-time applications. Within deep learning-based methods, Siamese U-Net has the fastest inference time at $90$ FPS. However, here we notice that all deep learning methods need to use GPU for real-time performance, while the TCF~\cite{ma2018novel} method can achieve $10$ FPS on a core i7 CPU. A visualization of the segmented results of all methods can be found in Fig.~\ref{Fig_result}. More qualitative results can be found in our supplemental video.

To conclude, our FW-Net can effectively learn the temporal continuity and significantly improves over the state of the art. Our method is also end-to-end and does not require data augmentation or any extra post-processing. The inference time of FW-Net is $15$ FPS on a GPU which allows it to be used in wide range clinical applications. More details about our project can be found at \url{https://sites.google.com/site/cathetersegmentation/}.

\section{Conclusions and Future Work}\label{Sec_con}
We propose FW-Net, an end-to-end and real-time deep learning framework for catheter and guidewire segmentation in 2D X-ray fluoroscopy sequences. Our FW-Net consists of three components to effectively learn the temporal information: a segmentation network, a flow network, and a novel flow-guided warping function. We showed that by learning the temporal continuity, the segmentation result can be improved even when training with the raw ground-truth data. The experimental results demonstrate that our FW-Net not only achieves state-of-the-art results, but also has real-time performance. Hence, the proposed approach can be integrated to robotic control frameworks or considered for generation of haptic feedback with deployment to various endovascular applications.

Since we use a vision-based method to automatically generate ground-truth with only binary segmentation mask, our FW-Net is currently tested with the binary segmentation problem. In the future, we would like to explore the ability of FW-Net in multiclass segmentation problem with X-ray images, where we can have more classes such as catheter, guidewire, blood vessel. This will allow FW-Net to become more useful in clinical scenarios. This further motivates application to closed-loop control with robotic platforms \cite{abdelaziz2019toward} that facilitate individual manipulation of catheters and guidewires. The proposed methodology will be prospectively fused with advanced user assistance to incorporate the entire interaction of endovascular instruments and vascular structures for adaptive generation of haptic feedback \cite{dagnino2018haptic}. Finally, the contribution bears great potential for integration into a novel skill assessment framework with image-based metrics in endovascular surgery. 

\section*{Acknowledgment}
\addcontentsline{toc}{section}{Acknowledgment}
We would like to thank A. Vlontzos for the useful discussion. This research is supported by the UK Engineering and Physical Science Research Council (EP/N024877/1) and the Wellcome Trust. 


\bibliographystyle{class/IEEEtran}
\bibliography{class/IEEEabrv,class/reference}
   
\end{document}